# Prediction of number of cases expected and estimation of the final size of coronavirus epidemic in India using the logistic model and genetic algorithm

GaneshKumar M, Soman.K.P, Gopalakrishnan.E.A, Vijay Krishna Menon, Sowmya.V

Center for Computational Engineering and Networking (CEN), Amrita School of Engineering, Coimbatore,
Amrita Vishwa Vidyapeetham, India.

mganesh1996kumar@gmail.com, kpsoman2000@gmail.com, ea_gopalakrishnan@cb.amrita.edu,
vijaykrishnamenon@gmail.com, v_sowmya@cb.amrita.edu

(March 2020)

## Abstract

In this paper, we have applied the logistic growth regression model and genetic algorithm to predict the number of coronavirus infected cases that can be expected in upcoming days in India and also estimated the final size and its peak time of the coronavirus epidemic in India.

## 1. Introduction

The epidemic of person-to-person transmissible pneumonia caused by the severe acute respiratory syndrome coronavirus 2 (SARS-COV-2, also known as COVID-19) has sparked a global alarm. The COVID-19 virus spreads primarily through droplets of saliva or discharge from the nose when an infected person coughs or sneezes. As of 23rd March 2020, there have been around 294,110 confirmed cases reported, with 12,944 deaths all around the world, affecting around 177 countries and territories. Milan Batista [1] used the logistic model in estimation of the final size and its peak time of the coronavirus epidemic in China, South Korea, and the rest of the World, which gave a reasonable description of the epidemic in those places. In this paper, we try to estimate the final size of the epidemic for India and also do daily predictions, that is predicting the expected number of infected cases in upcoming days using the logistic model and genetic algorithm with the data compiled by the Johns Hopkins University Center for Systems Science and Engineering (JHU CCSE) from various sources including the World Health Organization (WHO).

## 2. Estimations for India

Based On the available data as of 24[th] March 2020, the predicted final size of coronavirus epidemic in India using the logistic model and genetic algorithm is approximately 111185 cases and the peak of the epidemic will be on 18-Apr-2020. The estimated initial doubling time (time required for the number of cases to double) is 3.6 days.

Table 1. Estimated Final epidemic size for India

| Method | Final epidemic size (No of persons infected) |
|---|---|
| Logistic Growth Regression | 72766 |
| Logistic Growth Regression along with Genetic Algorithm search | 111185 |

Short term forecasting is given in Table 2, which gives the predictions made for number of cases expected in upcoming days in India. We can see that the error percentages for predictions of the last 3 days lies within 5%, but the error percentages of initial predictions are relatively high.

Table 2. Short-term forecasting for India

| Day | Date | Actual | Predicted | Error % |
|---|---|---|---|---|
| 49 | 18-Mar-2020 | 156 | 192 | 23.08 |
| 50 | 19-Mar-2020 | 194 | 230 | 18.56 |
| 51 | 20-Mar-2020 | 244 | 275 | 12.70 |
| 52 | 21-Mar-2020 | 330 | 330 | 0.00 |
| 53 | 22-Mar-2020 | 396 | 396 | 0.00 |
| 54 | 23-Mar-2020 | 499 | 475 | 4.81 |
| 55 | 24-Mar-2020 | - | 569 | |
| 56 | 25-Mar-2020 | - | 682 | |
| 57 | 26-Mar-2020 | - | 817 | |
| 58 | 27-Mar-2020 | - | 979 | |
| 59 | 28-Mar-2020 | - | 1172 | |
| 60 | 29-Mar-2020 | - | 1403 | |

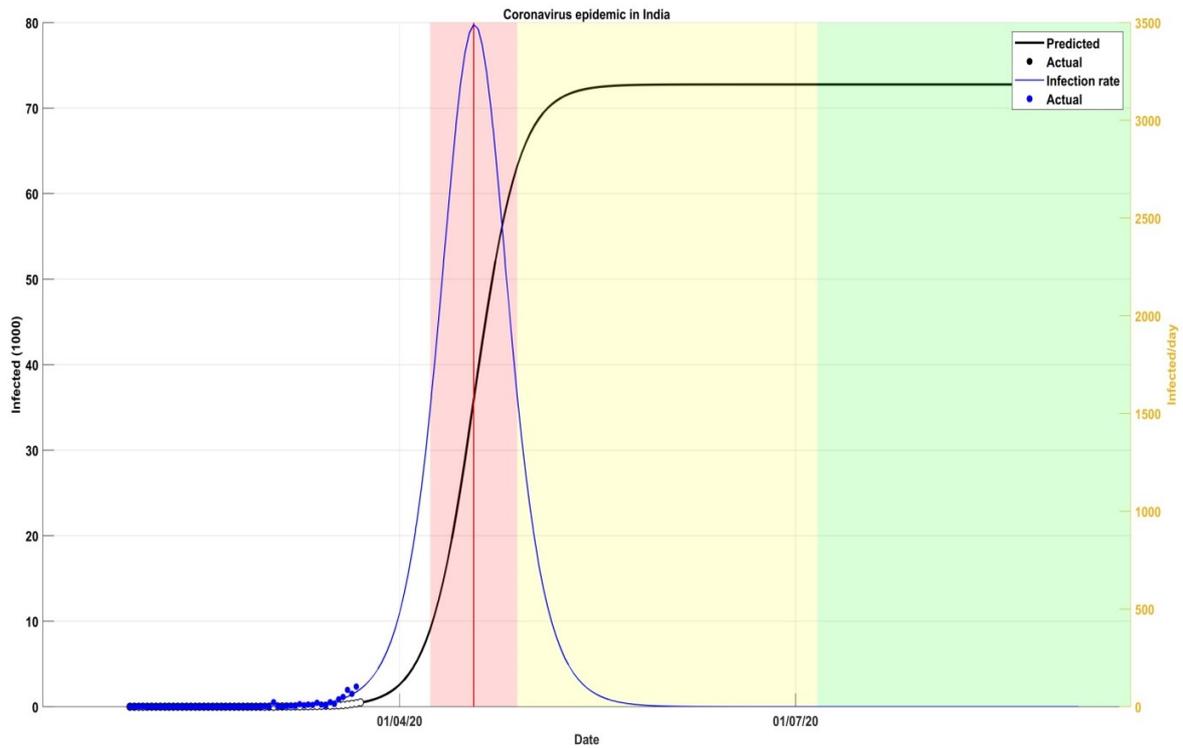

**Figure 1. Estimated logistic model for India using least square fit**

**Table 3. Estimated logistic model parameters for India using only least square fit**

| *K* | *r* | *A* |
|---|---|---|
| 72766 | 0.1919 | 3958053 |

**Table 4. Estimated logistic model parameters for India using least square fit and least absolute value fit using genetic algorithm**

| *K* | *r* | *A* |
|---|---|---|
| 111185 | 0.1820 | 3606403 |

## 3. Logistic growth model and Genetic algorithm

In the field of mathematical epidemiology, where we use mathematical models to study how infectious disease progress and show the likely outcome of an epidemic, one can use the Phenomenological model, which will describe the empirical relationship of different phenomena associated with the disease epidemic. Using the phenomenological approach, the epidemic dynamics can be described by the following variant of the logistic growth model [2-5]

$$\frac{dC}{dt} = rC\left(1 - \frac{C}{K}\right) \qquad (1)$$

Where C is an accumulated number of cases, r > 0 infection rate, and K > 0 is the final epidemic size. If $C(0) = C_0 > 0$ is the initial number of cases, then the solution of (1) is

$$C(t) = \frac{K}{1 + A\exp(-rt)} \qquad (2)$$

And we have the limiting condition $\lim_{t\to\infty} C(t) = K$ which (2) follows, where $K$ is the final size of the epidemic which has to be estimated.

There are three parameters in the logistic model (2): *K*, *r*, and *A* which should be calculated by least-square fit using the MATLAB function *lsqcurvefit*. Starting with initial guess of parameters $K_0$, $r_0$, and $A_0$, *lsqcurvefit* finds the final parameters $K_n$, $r_n$, and $A_n$ to best fit the nonlinear function (2) (solution of the logistic growth function) in accordance to our dataset. *lsqcurvefit* finds any optimal coefficient *x* in accordance to the given data by solving the following optimization problem:

$$\min_x \|F(x, xdata) - ydata\|_2^2 = \min_x \sum\left(F(x, xdata_i) - ydata_i\right)^2 \qquad (14)$$

Where *F(x, xdata)* is the prediction made by the nonlinear function for *xdata*, with respect to coefficient *x*. And *ydata* is the actual data.

The initial guess of the parameters $K_0$, $r_0$, and $A_0$ and the prediction of final epidemic size using iterated Shanks transformation [6], can be done following the similar procedure in the paper by Milan Batista [1].

Once we obtained the final parameters $K_n$, $r_n$, and $A_n$ using least-square fit, we can further re-estimate the parameters by doing a least absolute value fit, by applying Genetic algorithm search with a constraint that the values of the parameter can vary up to 10%. Genetic algorithm searches for the parameters which minimizes the absolute value of difference between actual and predicted values, obeying given the constraints. Genetic algorithm is available in Global optimization tool box in MATLAB.

## 4. Conclusion

Based On the available data as of $24^{th}$ March 2020, the predicted final size of coronavirus epidemic in India using the logistic model and genetic algorithm search is approximately 111185 cases. Regression convergence depends significantly on the initial guess of parameters *K*, *r* and *A*, and it may fail to converge for small data set or when initial guess of parameters are randomly assigned. We intend to further improve our model by collecting more data from upcoming days.

**Acknowledgements**
Code from the paper by Milan Batista [1] is used for obtaining the plots of logistic model.